\documentstyle[12pt]{dubna98}
\begin{document}
\begin{center}
{\bf THE DERIVATION OF DESER'S FORMULA AND ELECTROMAGNETIC CORRECTIONS
TO THE PIONIUM LIFE TIME\\}
\vspace*{1cm}
G. Rasche and A. Gashi\\
{\it Institut f\"{u}r Theoretische Physik \\}
{\it der Universit\"{a}t Z\"{u}rich,
Winterthurerstrasse 190,\\ CH-8057 Z\"{u}rich, Switzerland}
\end{center}

\vspace*{.5cm}
\begin{abstract}
{\small We give the modern derivation of Deser's formula using analytic continuation 
of the scattering amplitude as a function of momentum. The electromagnetic 
corrections to the pionium lifetime are given as evaluated in a potential model.}
\end{abstract}

\vspace*{1cm}
\underline{1. Introduction}

\vspace*{.3cm}
The Deser formula, first given in Ref.[1], relates the low energy scattering properties 
of strongly interacting charged particles to the hadronic properties 
of the atoms built from them. The main examples are:

$\pi^{-}p$ elastic and inelastic scattering and $\pi$-mesic hydrogen $(\pi^{-}p)$:

\[
\pi^-p\rightarrow\pi^-p,\; \pi^0n \; or \; \gamma n \qquad with \qquad (\pi^-p)\rightarrow \pi^0n \; or \; \gamma n
\]

$\pi^{-}\pi^{+}$ elastic and inelastic scattering and pionium $(\pi^{-}\pi^{+})$:

\[
\pi^-\pi^+\rightarrow\pi^-\pi^+,\; \pi^0\pi^0 \; or \; \gamma \gamma \qquad with \qquad (\pi^-\pi^+)\rightarrow \pi^0\pi^0 \; or \; \gamma \gamma
\]

To be more specific from the start we concentrate on the $\pi^{-}\pi^{+}$ system. 
Nearly all information comes from the s-wave state, so we treat only 
$l=0$ and do not indicate this by a specific label. 
The $\gamma\gamma$ channel does not play a decisive role, so we forget it 
for the moment. In principle it could be included in the analysis.

If we compare physical $\pi^{-}\pi^{+}$ scattering with pure Coulomb scattering of two 
oppositely charged particles we measure certain scattering lengths; 
if we compare the bound state $(\pi^{-}\pi^{+})$ with the pure Coulomb bound 
state of such an atom we measure an energy shift and a 
finite lifetime of the ground state.

The aim of the Deser formula then is to relate the scattering 
lengths to the shift and decay width of the ground state of the atom. 
For the two channel scattering system we have two scattering lengths $a_{cc}$ 
and $a_{0c}$ :
\[
{\pi^{-}\pi^+}^{\nearrow ^{\textstyle{\pi^-\pi^+}\qquad :a_{cc}} }_{\searrow _{\textstyle{\pi^0\pi^0}\qquad \; :a_{0c}} }
\]

We will give the precise definitions of these quantities later. 
It will turn out that the shift of the ground state is dominated by a 
term proportional to $a_{cc}$ and that the width for the decay
\[
(\pi^{-}\pi^{+})\rightarrow \pi^{0}\pi^{0}
\]
is dominated by a term proportional to $|a_{0c}|^{2}$.
To understand the main ideas for the derivation of this result it is 
convenient to treat a simplified system first: We assume that the 
$\pi^{0}$ does not exist. In this case only elastic scattering 
would be possible above threshold and pionium would be stable; this one 
channel problem will be treated in Sect.2. In Sect.3. we will then treat the 
realistic two channel case and in Sect.4. the problem of isospin conservation 
and electromagnetic corrections to this symmetry will be dealt with.

\vspace*{.3cm}

\underline{2. One channel case}

\vspace*{.3cm}

As already described we assume that $\pi^0$ does not exist. We only 
deal with elastic $\pi^{-}\pi^{+}$ scattering and the bound state, 
which in this case could not decay for energetic reasons. But due to the 
hadronic interaction the energy of the bound state will be 
shifted compared with the energy of the pure Coulomb ground state. 
So we face the problem of connecting the energy shift with the scattering 
length for $\pi^{-}\pi^{+}$ elastic scattering. The problem has been solved in a 
semiquantitative way in Ref.[1]. A systematic method which also allows the 
inclusion of corrections has been worked out on the basis of analytic 
continuation of the scattering amplitude in Ref.[2,3].
To this end we have to distinguish between 3 different scattering amplitudes:
$f^{(H)}$ ($H$ for hadronic), $f^{(Cb)}$ ($Cb$ for pure Coulomb) and $f^{(T)}$ ($T$ for total).
In the following $q$ is the magnitude of the momentum of either 
particle in the cm-frame; $m_c$ is the mass of the $\pi^{\pm}$.

We will now treat the different scattering amplitudes in a way which 
later allows an easy generalisation to the two channel case.

\vspace*{.3cm}
\underline{-purely hadronic scattering}
\vspace*{.3cm}

\begin{equation}
  	f^{(H)-1}(q)=K^{(H)-1}(q^2)-iq.
  	\label{eq1}
  \end{equation}
$K^{(H)}$ is a real function of $q^2$, analytic near $q^2=0$.

One can expand $K^{(H)-1}(q^2)$ in the following way:
\[
  	K^{(H)-1}(q^2)=a^{(H)-1}+\frac{1}{2}r_{eff}^{(H)}q^2+O(q^4).
  \]
$a^{(H)}$ is the hadronic scattering length, i.e. the scattering length 
with the Coulomb interaction neglected; $r_{eff}^{(H)}$ is the corresponding 
effective range.

From Eq.(1) we have:

\begin{equation}
  	f^{(H)}(q)=\frac{K^{(H)}(q^2)}{1-iqK^{(H)}(q^2)}.
  	\label{eq2}
  \end{equation}
A pole of $f^{(H)}$ on the positive imaginary 	axis of the complex $q$ - plane 
corresponds to a hadronically bound state. The only possibility for such a 
pole near $q=0$ can come from a zero of the denominator in Eq.(2).
In our case there is no such hadronically bound state.

\vspace*{.3cm}
\underline{-pure Coulomb scattering}

\vspace*{.3cm}

The corresponding scattering amplitude (partial wave amplitude for $l=0$) is 
known explicitly:

\[
  	f^{(Cb)}(q)=\frac{e^{2i\sigma_{0}}-1}{2iq},
  \]
with
\[
  	e^{2i\sigma_{0}}=\frac{\Gamma(1+i\eta)}{\Gamma(1-i\eta)},
  \]
  
  \[
  	\eta=-\frac{\alpha m_c}{2q}\equiv-\frac{1}{qa_B}.
  \]
Here $\alpha$ is the fine structure constant and $a_B$ is the 
Bohr radius of pionium.
From the well known analytic properties of the $\Gamma$-function one 	
finds the poles of $f^{(Cb)}(q)$ on the positive imaginary axis of the complex 
$q$ - plane at the positions
 
  \begin{equation}
  	q_{n}=\frac{i}{n}\frac{1}{a_B},       \qquad n=1,2,3,...
  \end{equation}
They correspond to the Coulomb bound states for $l=0$.

\vspace*{.3cm}
\underline{-Mixed scattering}

\vspace*{.3cm}

We assume that there is a short range hadronic interaction as well as the long 
range Coulomb interaction, which are additive in the Hamiltonian. The resulting 
total scattering amplitude then is

\[
  	f^{(T)}=f^{(Cb)}+e^{2i\sigma_{0}}f,
  	\label{eq5}
  \]
with
\begin{equation}
  	f^{-1}(q)=C_0^{-2}(\eta)\lgroup K^{-1}(q^2)-2q\eta h(\eta)\rgroup-iq,
  \end{equation}
where
\[
  	C_0^{2}=e^{-\pi\eta}\Gamma(1+i\eta)\Gamma(1-i\eta).
  \]
  
$h(\eta)$ also is an explicitly known function, related to the digamma function.
It is well known that $K(q^2)$ has the same analytic properties as $K^{(H)}(q^2)$
near $q^2=0$:

\[
  	K^{-1}(q^2)=a^{-1}+\frac{1}{2}r_{eff}q^2+O(q^4).
  \]

Here $a$ is the scattering length in the presence of the Coulomb 		
potential and $r_{eff}$ the corresponding effective range.
We then have

\[
  	f^{(T)}(q)=e^{2i\sigma_{0}}\{ \frac{1}{2iq}+
  	\frac{C_0^2(q)K(q^2)}{1-K(q^2)(2q\eta h(\eta)+iqC_0^2(\eta))} \}-\frac{1}{2iq}.
  \]
We are now looking for the poles of $f^{(T)}$ as a function of	$q$.
$e^{2i\sigma_0}$ has (as we know already) poles at $q_n$ (see Eq.(3)) for the 
pure Coulomb states. 
A careful analysis shows that the curly bracket vanishes at $q_n$, so the product 
does not have a pole. The only poles of $f^{(T)}$ therefore can come from 
the solutions of (note that $\eta$ depends on $q$)

\begin{equation}
  	1-K(q^2)(2q\eta h(\eta)+iqC^2_0(\eta))=0 \quad \mbox{ or equivalently} \qquad f^{-1}(q)=0.
  \end{equation}

These lie on the positive imaginary axis, near the pure Coulomb poles;
the difference in the position gives the energy shift $\Delta E$ due to the 
hadronic interaction. The result for the $1s$ state is:

\begin{equation}
  	\Delta E=\frac{4a}{a_B}(1+c_1\frac{a}{a_B}+...)E_1,
  	\label{eq6}
  \end{equation}
where $E_1$ is the pure Coulomb ground state energy corresponding to $q_1=i/a_{B}$.	

\vspace*{.3cm}
If one replaces $a$ by $a^{(H)}$ and neglects the second term in the parenthesis 
of Eq.(6) one finds (in our notation) the result of the semiquantitative treatment in Ref.[1].
The second term in the parenthesis is (since $a/a_B\ll1$ ) numerically 
negligible compared to the accuracy expected from experiments. It contains 
the higher order hadronic effets on the bound state. Furthermore $c_1$ depends on $r_{eff}$; 	
this takes into account the subthreshold effect, which means that $K(q^2)$ has to be continued 
to negative values of $q^2$.

The difference between $a$ and $a^{(H)}$ is well known from the comparison of 
pp-scattering with np-scattering in the 		
$l=0$, $S=0$, $T=1$ state of the two nucleon system. In this case $a^{(H)}$ (measured 
from np	scattering (no Coulomb force!)) is approximately $2.5a$ ($a$ being 
measured from pp scattering). Thus the difference exists and can in 
principle be measured. In the two nucleon system it is enhanced by a so called 
"virtual" bound state just above threshold. In the $\pi^-\pi^+$ system it is fortunately	
not as big as that and can be treated by perturbation theory. Since it is an 
electromagnetic effect we expect a difference of a few percent.

\vspace*{.5cm}

\underline{3. Two channel case}

\vspace*{.5cm}

In reality we have to treat the coupled two channel case 
(still neglecting the $\gamma\gamma$-channel):
\[
{\pi^-\pi^+}^{\nearrow^{\textstyle{\pi^-\pi^+}}}_{\searrow_{\textstyle{\pi^0\pi^0}}}
\]
Let $q$ be the magnitude of the momentum of each incident 
particle in the center of momentum frame, $q_0(q)$ the magnitude 
of the momentum of each outgoing $\pi^0$ in that frame 
($m_0$ is the mass of $\pi^0$):
\[
q^2_0+m^2_0=q^2+m^2_c
\]
Again we treat only s-wave scattering, which is justified at 
very small energies. The scattering amplitudes $f^{(\cdot)}$ now
become a symmetric 2x2 matrix:
\[
F^{(\cdot)}(q)=\left( \begin{array}{cc}
F^{(\cdot)}_{cc}(q) & F^{(\cdot)}_{c0}(q) \\ 
F^{(\cdot)}_{0c}(q)=F^{(\cdot)}_{c0}(q) & F^{(\cdot)}_{00}(q) \end{array} \right)
\]
The matrix index $c$ refers to the $\pi^-\pi^+$ channel and 0 refers to the $\pi^0\pi^0$ channel.
We will treat this problem first by neglecting the Coulomb interaction in the 
scattering problem (i.e. above threshold). But we keep the option of treating 
${m_c}\neq{m_0}$.
We consider the two cases:
\[
 {\rm I:\/}  \; m_0=m_c \qquad;\qquad {\rm II:\/}  \; m_0<m_c
\]
I: Since besides the Coulomb interaction we have also neglected the mass 
difference we are in the hadronic situation and indicate this by a 
superscript ($H$) in the scattering amplitude and scattering cross sections. 
We are dealing with coincident thresholds and therefore 
$F^{(H)}(0)=a^{(H)}$, with real matrix elements $a^{(H)}_{cc}$, 
$a^{(H)}_{c0} \! = \! a^{(H)}_{0c}$, $a^{(H)}_{00}$ as in the single 
channel system. The limiting values of the $\pi^-\pi^+$ elastic and charge 
exchange  cross sections are:
\[
a) \:	\lim_{q\rightarrow 0}\sigma ^{(H)}_{cc}(q)=4 \pi | F^{(H)}_{cc}(0)|^2=4\pi|a^{(H)}_{cc}|^2 \\
\]
\[
b) \:	\lim_{q\rightarrow 0}\sigma ^{(H)}_{0c}(q)=4\pi|F^{(H)}_{0c}(0)|^2=4\pi|a^{(H)}_{0c}|^2
\]
II: Here the mass difference is taken into account and we indicate this by a 
superscript $(\Delta)$.
The matrix $F^{(\Delta)}{(0)}$ now has complex matrix elements. The limiting 
behaviour of the cross sections is:
\[
a) \:	\lim_{q\rightarrow 0}\sigma^{(\Delta)}_{cc}(q)=4\pi|F^{(\Delta)}_{cc}(0)|^2 \\
\]
\[
b) \:	\lim_{q\rightarrow 0}\sigma^{(\Delta)}_{0c}(q) \sim 4 \pi \frac{q_{0}(0)}{q}|F^{(\Delta)}_{0c}(0)|^2
\]
Going over from I$a$ to the more realistic situation II$a$ is not a dramatic change. 
We can expect that $F^{(\Delta)}_{cc}(0)$ and $a^{(H)}_{cc}$ differ by only a few percent. 
But going over from I$b$ to II$b$ changes the low energy behaviour of 
$\sigma^{(\Delta)}_{0c}(q)$ drastically, because we are now dealing with a real 
exothermic reaction. Nevertheless we can expect that 
$F^{(\Delta)}_{0c}(0)$ and $a^{(H)}_{0c}$ also differ by only a few percent. Going from I to II 
obviously breaks a possible isospin invariance of the hadronic interaction.

We now treat case II more systematically. As in the single channel case we 
have
\begin{equation}
F^{(\Delta)-1}(q)=K^{(\Delta)-1}(q^2)-iQ(q),
\end{equation}
where
\[
Q(q)=\left( \begin{array}{cc} q & 0 \\ 0 & q_0(q) \end{array} \right) 
, \qquad
Q(0)=\left( \begin{array}{cc} 0 & 0 \\ 0 & q_0(0) \end{array} \right) 
\]
From unitarity one can show that $K^{(\Delta)}(q^2)$ is a real symmetric 
2x2 matrix and that in analogy with the single channel case
\[
K^{(\Delta)-1}(q^2)=A^{(\Delta)-1}+O(q^2)
\]
For our present pedagogical purpose we only need Eq.(7) at $q=0$:
\[
F^{(\Delta)-1}(0)=A^{(\Delta)-1}-iQ(0)
\]
Solving for the matrix elements of $F^{(\Delta)}(0)$ we get:
\[
F^{(\Delta)}_{cc}(0)=A^{(\Delta)}_{cc}+iq_{0}(0)\frac{(A^{(\Delta)}_{0c})^2}{1-iq_{0}(0)A^{(\Delta)}_{00}} \\
\]
\[
F^{(\Delta)}_{0c}(0)=\frac{A^{(\Delta)}_{0c}}{1-iq_{0}(0)A^{(\Delta)}_{00}}.
\]
In the limit $m_{0}=m_{c}$ we have $q_{0}(0)=0$ and $F^{(\Delta)}(0)=A^{(\Delta)}(=a^{(H)})$.

We see that, in contrast to the one channel case, the threshold elastic scattering
amplitude $F^{(\Delta)}_{cc}(0)$ now has an imaginary part. Replacing $a$ in 
Eq.(6) by $F^{(\Delta)}_{cc}(0)$, we have (neglecting terms of higher order in the mass difference
 and $O(a/a_B)$):
\begin{equation}
\Delta E^{(\Delta)}=\frac{4A^{(\Delta)}_{cc}}{a_B}E_1+i4q_{0}(0)
\left(A^{(\Delta)}_{0c}\right)^2\frac{E_1}{a_B} 
\end{equation}
Here the real part of $\Delta E^{(\Delta)}$ gives the energy shift of 
the ground state as found before. The imaginary part of $\Delta E^{(\Delta)}$ 
is connected to the line width $\Gamma$ and the life time $\tau$ by
\[
Im\Delta E^{(\Delta)}=-\frac{1}{2}\Gamma^{(\Delta)};  \qquad \tau^{(\Delta)}=\Gamma^{(\Delta)-1}
\]
We therefore have
\begin{equation}
\tau^{(\Delta)-1}=-8q_{0}(0)\left(A^{(\Delta)}_{0c}\right)^2\frac{E_1}{a_B}.
\end{equation}
This phenomenological derivation of Deser's formula for the two channel case 
has been given in Ref.[4].

The final systematic derivation of Deser's formula has been achieved in Ref.[5].
It includes the Coulomb interaction also above threshold in addition 
to the mass differences and makes use of the method of analytic continuation 
in the two channel case. In analogy with Eq.(4) a matrix $F(q)$ 
can be defined by
\begin{equation}
F^{-1}(q)=K^{-1}(q^2)-iQ(q)+\;Coulomb \;terms \;in \;the \;charged \;channel,
\end{equation}
where $K(q^2)$ again is a real symmetric matrix analytic near $q^2=0$. 
We define $K(0)=A$. We do not give the Coulomb terms 
explicitly nor do we comment on the analytic continuation as we did in Sect.2. 
The result is that we have (analogously to Eq.(5)) to look for the zeros 
of $detF^{-1}(q)$. These zeros are off the positive imaginary $q$-axis, 
corresponding to complex energy shifts of the Coulomb poles as in Eq.(8).
The interpretation is the same as given after Eq.(8).
We only give the full result for the lifetime $\tau$ of the ground state:
\begin{equation}
\frac{1}{\tau}=-\frac{8}{a_{B}}\frac{\bar{q}_{0}(A_{0c})^2}
{1+{\bar{q}_{0}}^2(A_{00})^2}E_1\left(1+O
\left( \frac{A_{0c}}{a_B}\right)+\;subthreshold \;corrections \right)
\end{equation}
Here $\bar{q}_0$ is the magnitude of the momentum of each $\pi^0$ in the 
rest frame of the pionium atom.
Neglecting the numerically unimportant subthreshold corrections as well as the 
terms of $O({A_{0c}}/{a_B})$ and 
$O\left({\bar{q}_{0}}^2(A_{00})^2\right)$ we get nearly back to 
Eq.(9); but note the difference between 
$\bar{q}_0$ and $q_{0}(0)$ 
and the different definitions of $A^{(\Delta)}_{0c}$ and $A_{0c}$.

We emphasize very strongly that the Deser formula connects measurable 
quantities with measurable quantities: 
The matrix elements of $A$ can "in principle" be measured in 
low energy $\pi^-\pi^+$ elastic and inelastic scattering and $\pi^0\pi^0$ 
elastic scattering. The expressions for the cross sections in terms of 
$A_{cc}$, $A_{0c}$, $A_{00}$, including mass differences and the Coulomb effects 
near threshold are given in Ref.[6] for the $\pi^-p$ - system and can easily be 
translated to the $\pi^-\pi^+$ - system.

\vspace*{.5cm}

\underline{4. Isospin invariance and electromagnetic corrections to the 
pionium life time}

\vspace*{.5cm}

Up to now we have not made any use of a specific model for the hadronic 
interaction nor did we assume a symmetry property like isospin invariance.

From now on we use the well known result of chiral perturbation theory that the 
hadronic interaction in the $\pi\pi$ - system is isospin invariant. 
Whilst the proper understanding of radiative corrections to this result needs 
a calculation in the framework of chiral perturbation theory Ref.[7], an  
estimate of their order of magnitude can certainly  be achieved in a simple 
phenomenological model. Such a model has been developed in Ref.[8,9] for the 
$\pi^-\pi^+$ - system. It assumes that the bulk of the electromagnetic effects 
has its origin in the mass difference between $\pi^-$ and $\pi^0$ and in the 
Coulomb interaction between $\pi^-$ and $\pi^+$. This method has been applied 
in the past to the $pp$ - system and to the $\pi^-p$ - system (Ref.[5,6]).

We will concentrate here on the application to the lifetime of pionium; so 
we are interested in the quantity

\begin{equation}
	A_{0c} - a^{(H)}_{0c},
\end{equation}
with
\begin{equation}
	a^{(H)}_{0c} = \frac{\sqrt{2}}{3}(a^2-a^0),
\end{equation}
where $a^2$ and $a^0$ are the hadronic values of the s-wave scattering lengths
for total isospin 2 and 0 respectively. These two quantities come from chiral 
perturbation theory without electromagnetic interactions and can be evaluated 
either for the reference mass $m_0$ or for the reference mass $m_c$.

It is well known from chiral perturbation theory that the hadronic mass of 
the pions is practically equal to $m_0$ and that the difference $m_c-m_0$ is 
of electromagnetic origin. Therefore taking $m_0$ as the reference mass for 
evaluating $a^{(H)}_{0c}$ does result practically in a purely hadronic value.
On the other hand, taking $m_c$ as the reference mass results in an isospin 
invariant, but not purely hadronic, value for $a^{(H)}_{0c}$. Neverthless, 
as in most of the chiral perturbation theory work, we take $m_c$ as the 
reference mass for the following.

We emphasize also that we calculate the electromagnetic 
\underline{corrections} to the hadronic amplitudes and that the values of the 
corrections are hardly affected at all by the uncertainty in the starting 
values of $a^2$ and $a^0$ (one loop result, two loop result, generalized 
chiral perturbation theory, ...).

We now describe how the potential model is used to estimate the electromagnetic 
corrections to the scattering lengths for physical processes. Of special 
interest is the difference in Eq.(12).
Our input is the isospin invariant hadronic scattering phase shifts 
 $\delta^2(q)$  and  $\delta^0(q)$ given by chiral perturbation theory with 
the reference mass $m_c$. The isospin invariant scattering lengths are 
then given by 
\[
a^T = \lim_{q\rightarrow 0}\left(\delta^T(q)/q\right) \quad T=0, 2
\]
Taking as an example the results of Ref.[10] we get:
\[
	a^0=0.2883 \; {\rm fm\/}, \quad a^2=-0.0617 \; {\rm fm\/}
\]
and from Eq.(12)
\[
	a^{(H)}_{0c}=-0.1650 \; {\rm fm.\/}
\]
In each isospin channel  $T$ we then construct a potential $V^T$  
which inserted in a relativistically modified Schrodinger equation with 
reduced  mass $\frac{1}{2}m_c$  reproduces these phase shifts and 
scattering lengths.

The two potentials  $V^T$ are then used in a relativistically
modified coupled two channel Schrodinger equation in which the 
Coulomb potential is added in the charged channel and the 
physical masses $m_c$, $m_0$ are used.
Solving this two channel scattering problem we get the matrix 
 $K(q^2)$  (see Eq.(10)) and  $A=K(0)$. 
Thus we have the matrix element  $A_{0c}$  which dominates the 
Deser formula Eq.(11) for the lifetime of pionium. 
The result of this calculation is
\begin{equation}
	A_{0c}-a^{(H)}_{0c}=0.005(1) {\rm fm\/}
\end{equation}
The $20 \%$ error on the electromagnetic correction in Eq.(14) reflects the 
fact that the construction of an energy independent  $V^{T}(r)$  from 
 $\delta^T(q)$  is not unique. We have tested several forms for 
such a potential which reproduce the phases well up to 500 MeV cm energy.
We have also tested energy dependent potentials  $V^T$. In all these cases, 
the difference in Eq.(14) lies within the error given.

Numerically we arrive at the following relation connecting the hadronic quantity 
$a^{(H)}_{0c}$ given in Eq.(13) and the lifetime of pionium 
$\tau$ :
\[
	\tau({\rm fs\/})=\frac{0.0951}{\left(a^{(H)}_{0c}({\rm fm\/})+0.005\right)^2}
\]
\begin{center}
	{($a^{(H)}_{0c}$  evaluated using the reference  mass $m_c$).}
\end{center}

We have emphasised explicitly once more that $m_c$  has been taken 
as the reference mass. If $m_0$ is used instead, the potentials 
$V^T$  have to be determined from a Schrodinger equation with a reduced 
mass $\frac{1}{2}m_0$  and the electromagnetic correction has a 
different value. This point of view was taken in Ref.[8], where the 
corresponding results can be found.

We also emphasise that the influence of the $\gamma\gamma$ channel on the electromagnetic corrections 
to the scattering lengths can be incorporated by doing a three channel analysis.
The analogous treatment of the $\gamma n$ -channel in the $\pi^- p$ -system has been performed 
in Ref.[5]. Numerically this additional correction is negligible.

We thank W.S. Woolcock for a very careful reading of the manuscript.


\vspace*{.5cm}
\begin{thebibliography}\\
\bibitem{a1} S.Deser et al., {\it Phys.Rev.} {\bf 96} (1954) 774.
\bibitem{a2} T.L.Trueman, {\it Nucl.Phys.} {\bf 26} (1961) 57.
\bibitem{a3} E.Lambert, {\it Helv.Phys.Acta} {\bf 42} (1969) 667.
\bibitem{a4} H.Pilkuhn and S.Wycech, {\it Phys.Lett.} {\bf 76B} (1978) 29.
\bibitem{a5} G.Rasche and W.S.Woolcock, {\it Nucl.Phys.} {\bf A381} (1982) 405.
\bibitem{a6} G.Rasche and W.S.Woolcock, {\it Helv.Phys. Acta} {\bf 49} (1976) 455.
\bibitem{a7} V.Lyubovitskij and A.Rusetsky, {\it Phys. Lett} {\bf B389} (1996) 181, 
K.Maltman and C.E.Wolfe, {\it Phys. Lett.} {\bf B393} (1997) 19,  
U.G.Meissner et al., {\it Phys. Lett.} {\bf B406} (1997) 154,
Erratum:{\it Phys. Lett.} {\bf B407} (1997) 454, 
 V.Lyubovitskij et al., {\it JETP Lett.} {\bf 66} (1997) 783. 
 See also the contributions of A.G.Rusetsky, H.Sazdjian and E.A.Kuraev in these proceedings.
\bibitem{a8} U.Moor et al., {\it Nucl.Phys.} {\bf A587} (1995) 747.
\bibitem{a9} A.Gashi et al., {\it Nucl.Phys.} {\bf A628} (1998) 101.
\bibitem{a10} J.Gasser and H.Leutwyler, {\it Ann. of Phys.} {\bf 158} (1984) 142.


\end{thebibliography}
\end{document}